\documentclass{elsart}
\usepackage{natbib}
\input psfig.sty
\input epsfig.sty
\begin{document}
\rightline{\vbox{\halign{&#\hfil\cr
SLAC-PUB-8544\cr
August 2000\cr}}}
\vspace{0.8in} 
\def\ie{{\it i.e.}}
\def\eg{{\it e.g.}}
\def\etc{{\it etc}}
\def\etal{{\it et al.}}
\def\ibid{{\it ibid}.}
\def\nn{\noindent}
\def\to{\rightarrow}
\runauthor{Rizzo}
\begin{frontmatter}
\title{New Physics Beyond the Standard Model at $\gamma \gamma$ Colliders
\thanksref{X}}
\author[SLAC]{Thomas G. Rizzo}
\thanks[X]{Work supported by the Department of Energy, Contract 
DE-AC03-76SF00515}

\address[SLAC]{Stanford Linear Accelerator Center, Stanford CA 94309, USA}
\begin{abstract}
The complementarity of $e^+e^-$ and $\gamma \gamma$ colliders to discover and 
explore new physics beyond the Standard Model(SM) is discussed. After briefly  
surveying a number of various new physics scenarios we concentrate in detail 
on signatures for Large Extra Dimensions via the process $\gamma \gamma \to 
WW$.
\end{abstract}
\end{frontmatter}

\typeout{SET RUN AUTHOR to \@runauthor}

\section{Introduction}

As is well known there are two ways in which new physics beyond the SM may 
manifest itself. In the direct scenario, a new particle (or set of 
particles) may be singly or pair produced at a collider. In the indirect 
scenario,  precision measurements would determine at a high level of 
confidence that deviations from the SM are found for a 
particular observable or set of observables. This set of deviations may or may 
not be sufficient in their magnitude or direction to point to the correct new 
physics scenario from which they arose. 

In the case of direct production it is important to note that the cross 
sections for pairs of scalars, fermions or vectors particles are all 
significantly larger at 
$\gamma \gamma$ colliders than they are at $e^+e^-$ colliders (even after 
angular acceptance cuts) with comparable luminosities anticipated at both 
machines. However, by comparison,  the $\gamma \gamma$ search reach is reduced 
due to the high energy cutoff in the 
photon fluxes. While polarized $\gamma \gamma$ colliders allow one to isolate 
the couplings of new physics to photons and probe spin configurations not 
accessible in $e^+e^-$ collisions, they cannot produce neutral particles 
except via loops, though such cross sections may in some cases be large. 
For indirect searches both polarized $e^+e^-$ and $\gamma \gamma$ 
colliders lead to a similar number of final states which have somewhat 
comparable statistical power as far as looking for deviations are 
concerned. However, in some cases the $\gamma \gamma$ collider 
has the edge due to the larger cross 
sections. Based on these arguments alone, and not knowing the source of new 
physics {\it a priori},  we would expect a general complementarity for 
$e^+e^-$ and $\gamma \gamma$ for new physics searches. Unfortunately, analyses 
of new physics scenarios are not yet as evolved in $\gamma \gamma$ collisions 
as they are for $e^+e^-$.

\section{Some Examples}

There are a number of examples of new physics scenarios which clearly display 
this complementarity. Perhaps the most well known amongst these possibilities 
are the anomalous gauge couplings of the $W$ and searches for leptoquarks. 
While there have been extensive analyses{\cite {rev}} of anomalous triple 
gauge boson couplings at $e^+e^-$ colliders, the corresponding processes 
$\gamma e \to W\nu$ and 
$\gamma \gamma \to WW$ have generally gotten less attention in the 
literature{\cite {old}}. Note that while $\gamma e \to W\nu$ and 
$\gamma \gamma \to WW$ isolate the anomalous photon couplings to the $W$,  
$e^+e^- \to WW$ also potentially involves anomalous $Z$ couplings 
which may also be present. This process is also dominated by the large 
$t$-channel neutrino exchange diagram which can be removed using beam 
polarization. Perhaps the best example of this complementarity is displayed 
in the study of Choi and Shrempp that compares the sensitivities of these 
three processes to the anomalous couplings 
$\Delta \kappa_\gamma=1-\kappa_\gamma$ and $\lambda_\gamma$ which is shown in 
Fig.1. Though now outdated, one sees from this analysis 
that the sensitivities of the three experiments are qualitatively 
similar and their overlapping region is substantially 
smaller than that obtained from any single measurement. It would be very 
interesting, and perhaps quite important, to 
repeat this analysis using modern luminosities which are more than an order 
of magnitude larger than used in this study and to include the increased set of 
observables that can be used to constrain these anomalous couplings. It is 
easy to imagine that the resulting area of overlap would now be smaller by 
more than two orders of magnitude. In the case of anomalous quartic 
gauge couplings, they can be probed at $\gamma \gamma$ colliders through 
processes which 
occur at order $\alpha^2$ whereas they can only be accessed in $e^+e^-$ 
collisions at order $\alpha^3$ giving a great advantage to $\gamma \gamma$ 
colliders. 

\begin{figure*}
\centerline{
\epsfig{figure=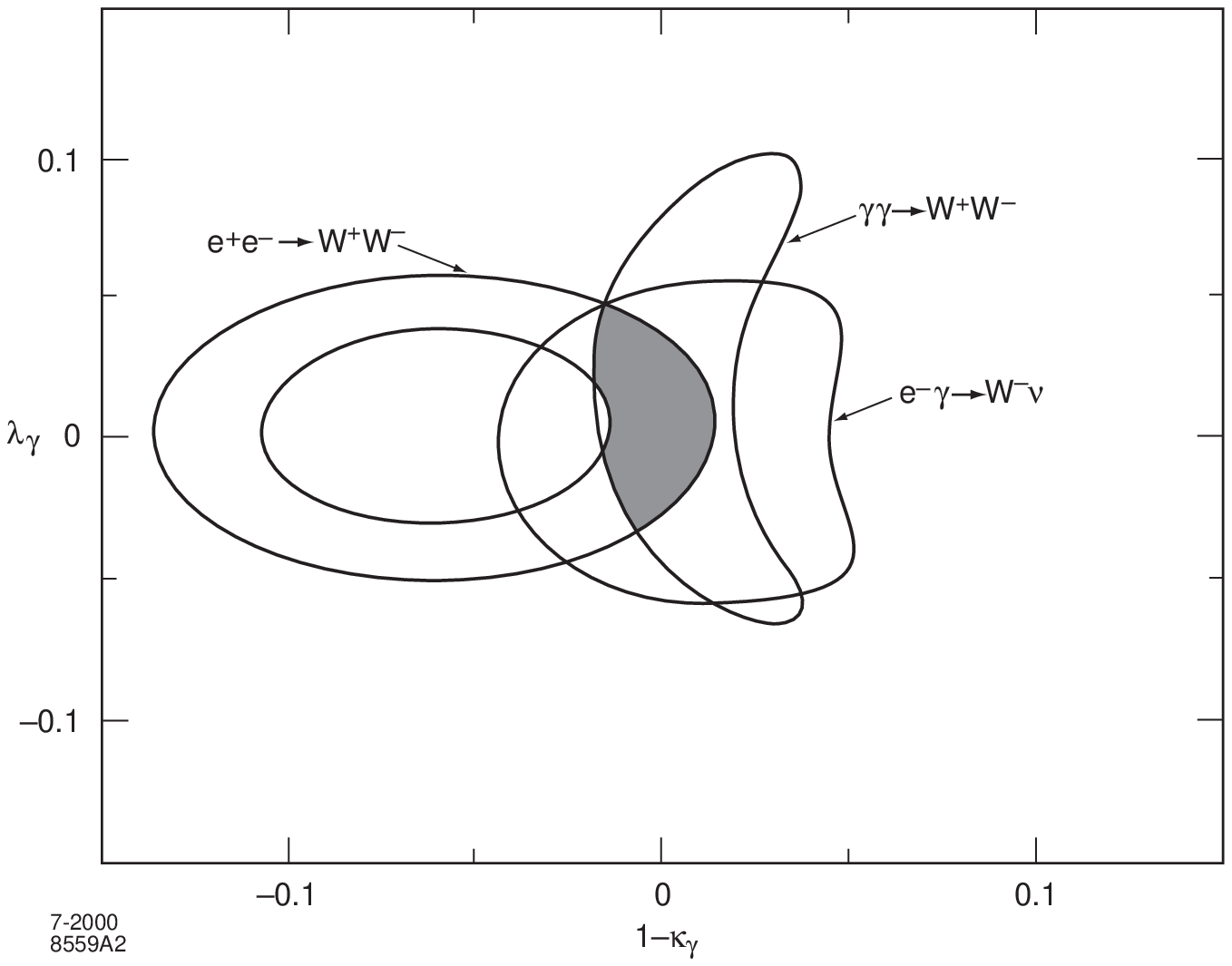,height=8cm,width=12cm,angle=0}}
\vspace*{0.1cm}
\caption[*]{Allowed overlapping regions from the analysis of Choi and Schrempp 
in the $\Delta \kappa_\gamma-\lambda_\gamma$ anomalous coupling plane.}
\end{figure*}
\vspace*{0.6mm}

Leptoquarks{\cite {lq}} have been well studied in $e^+e^-$, $\gamma e$ and 
$\gamma \gamma$ colliders. Buchm\" uller, R\" uckl and Wyler have long ago 
shown that under reasonably general assumptions leptoquarks can arise in any 
of 14 different varieties. Leptoquarks may have fermion number 
$F=0,\pm 2$, come 
in singlets, doublets or triplets of weak isospin and are either scalars or 
vectors. (There may even be some evidence for the existence of leptoquarks in 
the mass range near 400 GeV{\cite {zarn}}.) Since the LHC will run before any 
high energy $e^+e^-$ or $\gamma \gamma$ collider and discover(or not) any 
leptoquarks in the mass range accessible to these machines, their true role 
is not leptoquark discovery but leptoquark {\it identification}. 
Assuming these particles are elementary, a given leptoquark species 
is easily identified in $e^+e^-$ collisions through the energy dependence of 
it's total cross section and angular distribution. However, if form factors are 
present and/or leptoquarks have gauge strength Yukawa couplings it is possible 
that confusion can arise and a polarized 
$\gamma \gamma$ collider will be needed to disentangle the various ID 
possibilities. Since leptoquark charges, $Q$, vary in magnitude between $1/3$ 
and $5/3$ and their production cross sections at $\gamma \gamma$ are 
proportional to $Q^4$, a wide range of rates can be expected. This is shown 
explicitly in Fig.2 from {\cite {bel}} for a $\sqrt s$=1 TeV collider.  

Note that similar complementarity between $e^+e^-$ and $\gamma \gamma$ 
colliders also arises in the case of excited fermion production{\cite {stu}}.

\begin{figure*}
\centerline{\psfig{figure=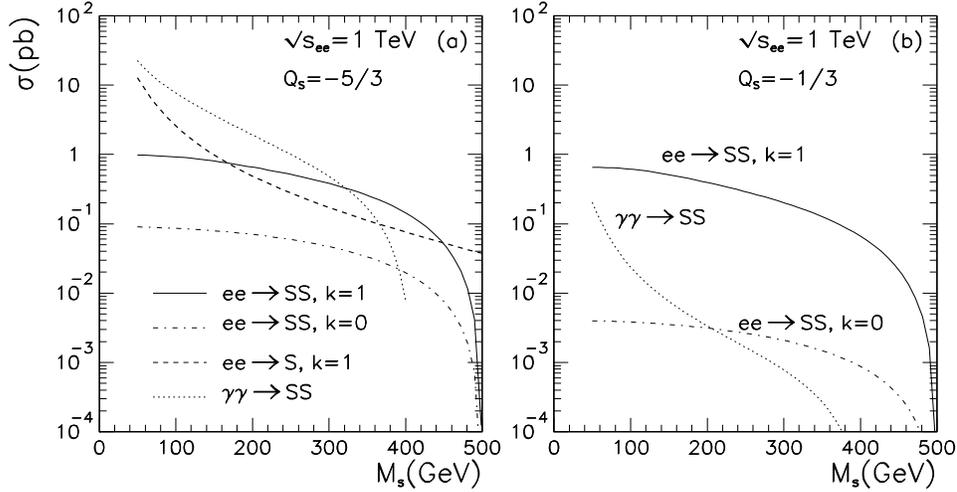,height=8cm,width=14cm}}
\vskip-0.6truecm
\caption{LQ pair-production cross sections at $e^+e^-$ and $\gamma\gamma$
colliders. Here $k$ is the Yukawa coupling strength in units of strength of 
$e$ for $t$- or $u$-channel quark exchange graphs. Cross sections are shown 
for the two extreme charge cases of $Q=5/3$ and $Q=-1/3$.} 
\end{figure*}

\section{$\gamma \gamma \to WW$ in Theories with Large Extra Dimensions}

In theories with large extra dimensions{\cite {add}} the exchange of 
Kaluza-Klein(KK) graviton towers result in large set of new dimension-8 
operators that can lead to new contributions 
to the matrix elements for a number of different processes{\cite {pheno}}. 
These exchanges can cause substantial deviations in cross sections, angular 
distributions and asymmetries. These operators in 
principle have an arbitrary sign, $\lambda=\pm 1$, and an associated cutoff 
mass scale, $M_H$, 
which is of order the higher dimensional Planck scale. Of the many processes 
examined so far, $\gamma \gamma \to WW$ provides the greatest reach for $M_H$ 
in comparison to the overall collider center of mass energy. (It is difficult 
to compare the $\sqrt s$ dependence of this reach for hadron colliders and 
$e^+e^-$/$\gamma \gamma$ colliders.) 
The main reasons for this are that the $WW$ final state offers many observables 
which are particularly sensitive to the initial electron and laser 
polarizations as well as the 
very high statistics available (due to the 80 pb cross section) with which 
to probe graviton KK contributions. The differential cross sections are shown 
in Fig.3 for the SM as well as when the KK tower is included for $\lambda=1$.

\begin{figure*}
\centerline{
\psfig{figure=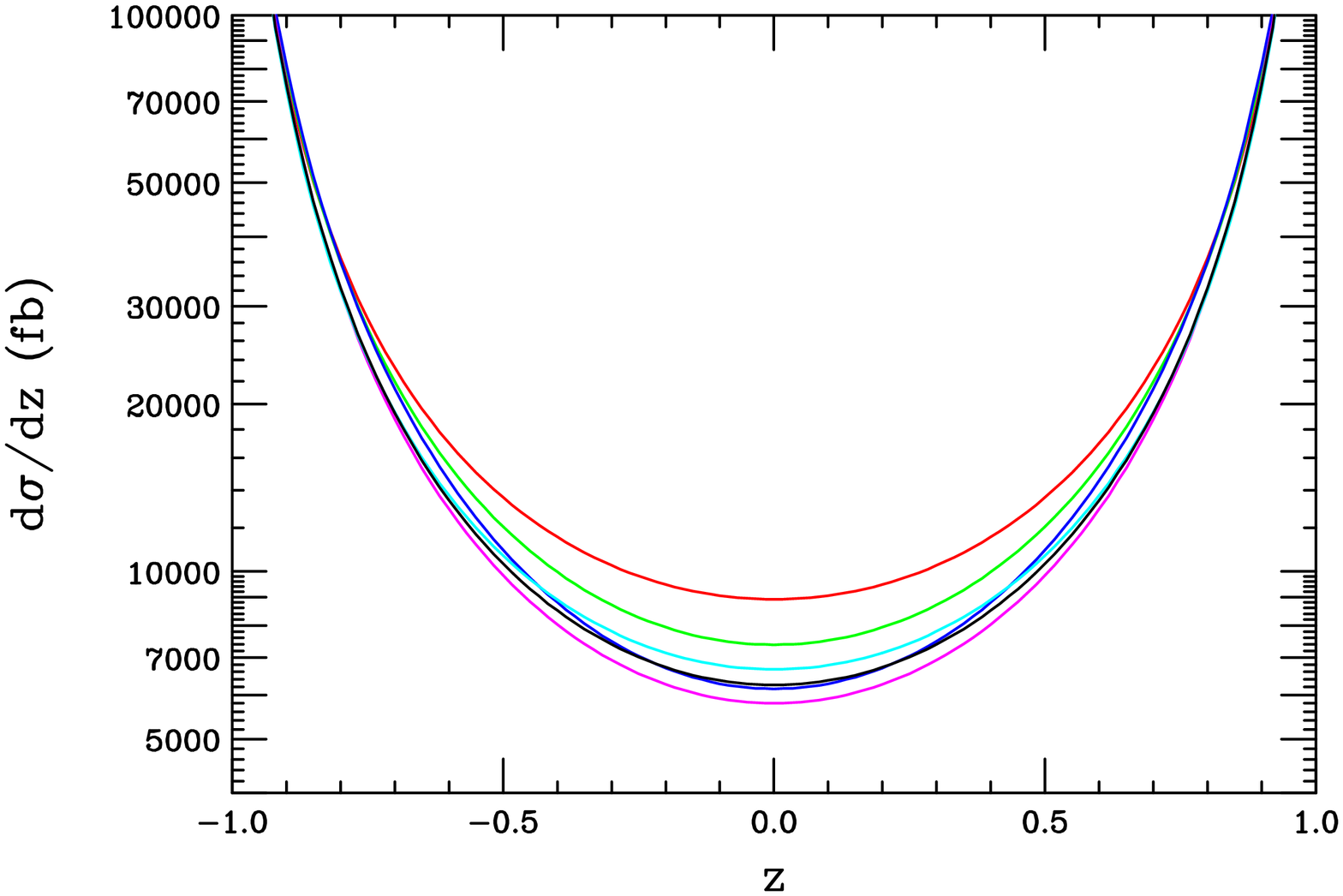,height=5.5cm,width=7cm,angle=90}
\hspace{-0.10cm}
\psfig{figure=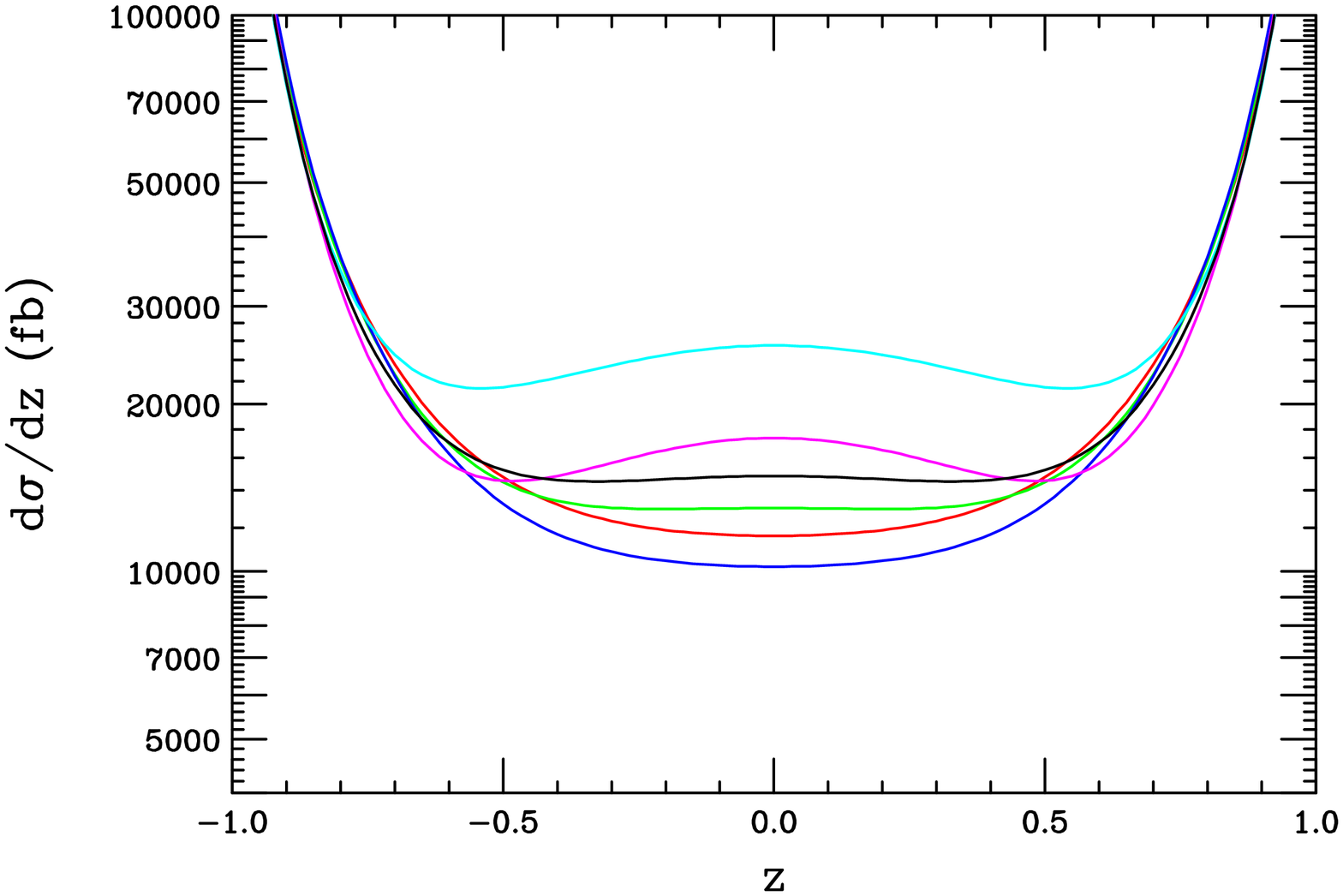,height=5.5cm,width=7cm,angle=90}}
\vspace*{0.1cm}
\caption{Differential cross section for $\gamma\gamma \to W^+W^-$ at a 1 TeV 
$e^+e^-$ collider for (left)the SM and with $M_H=2.5$ TeV with 
(right)$\lambda=1$. The $\lambda=-1$ results are quite similar. 
In (left) from top to bottom in the center of the figure 
the helicities are $(++++)$, 
$(+++-)$, $(-++-)$, $(++--)$, $(+---)$, and $(+-+-)$; in (right) they are  
$(-++-)$, $(+-+-)$, $(+++-)$, $(+---)$, $(++++)$, and $(++--)$, 
where we have employed the notation $(P_{e1},P_{l1},P_{e2},P_{l2})$.}
\end{figure*}
\vspace{3mm}

Note that within the SM there is no 
dramatically strong sensitivity to the initial state lepton and laser 
polarizations and all of the curves have roughly the same shape. 
When the graviton tower contributions are included there are several 
effects. First, all 
of differential distributions become somewhat more shallow at large 
scattering angles but there is little change in the forward and backward 
directions due to the dominance of the SM poles. Second, there is now a 
clear and distinct sensitivity to the initial state 
polarization selections. In some cases, particularly for the $(-++-)$ and 
$(+-+-)$ helicity choices, the differential cross section increases 
significantly for angles near $90^o$ taking on an m-like shape. This shape is, 
in fact, symptomatic of the spin-2 nature of the K-K graviton tower exchange 
since a spin-0 exchange leads only to a flattened distribution. 
Given the very large statistics available with a typical integrated luminosity 
of 100-300 $fb^{-1}$, it is clear that the $\gamma \gamma \to W^+W^-$ 
differential cross 
section is quite sensitive to $M_H$ especially for the two initial state 
helicities specified above. 

\begin{figure*}
\centerline{
\psfig{figure=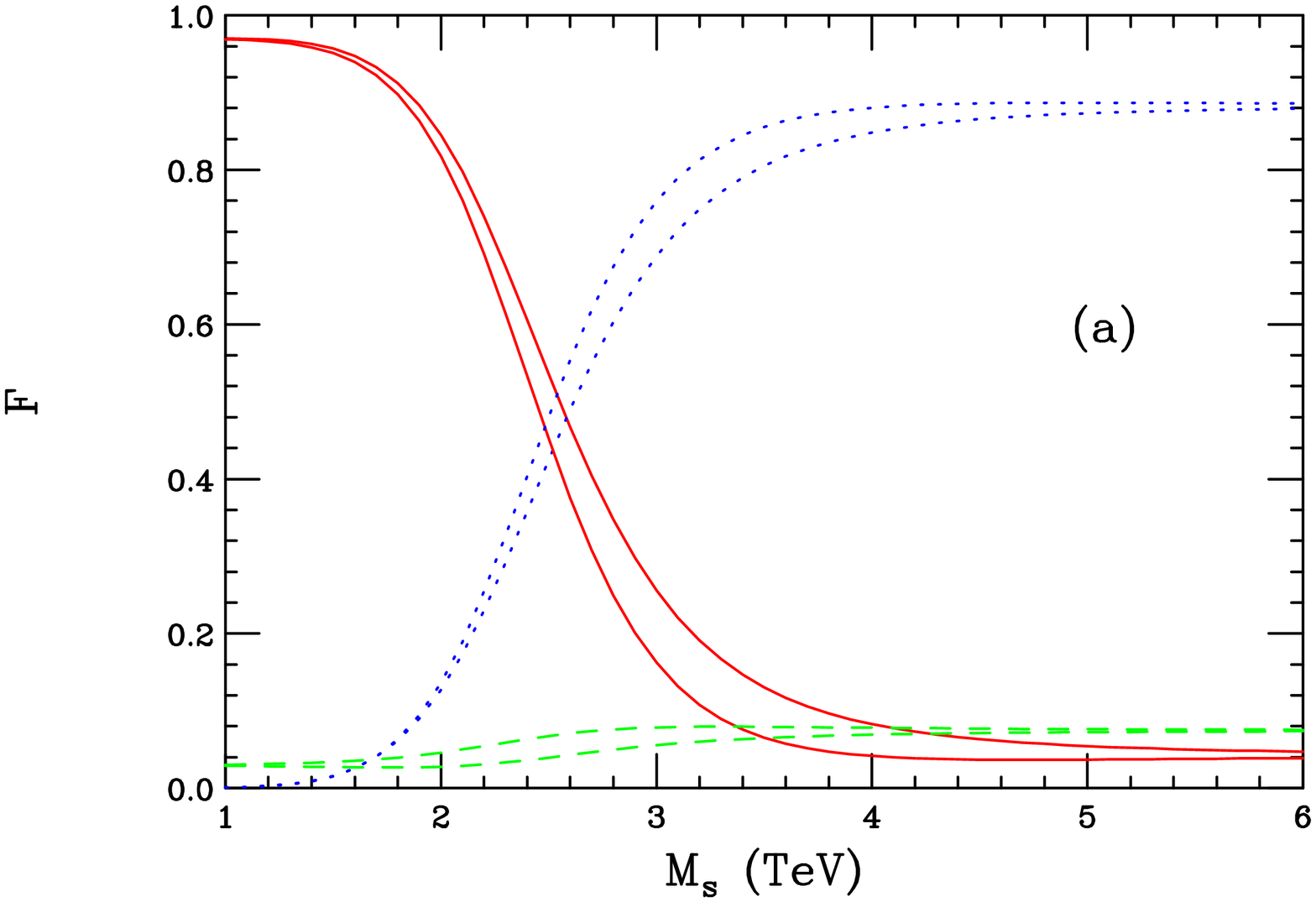,height=5.5cm,width=7cm,angle=90}
\hspace{-0.10cm}
\psfig{figure=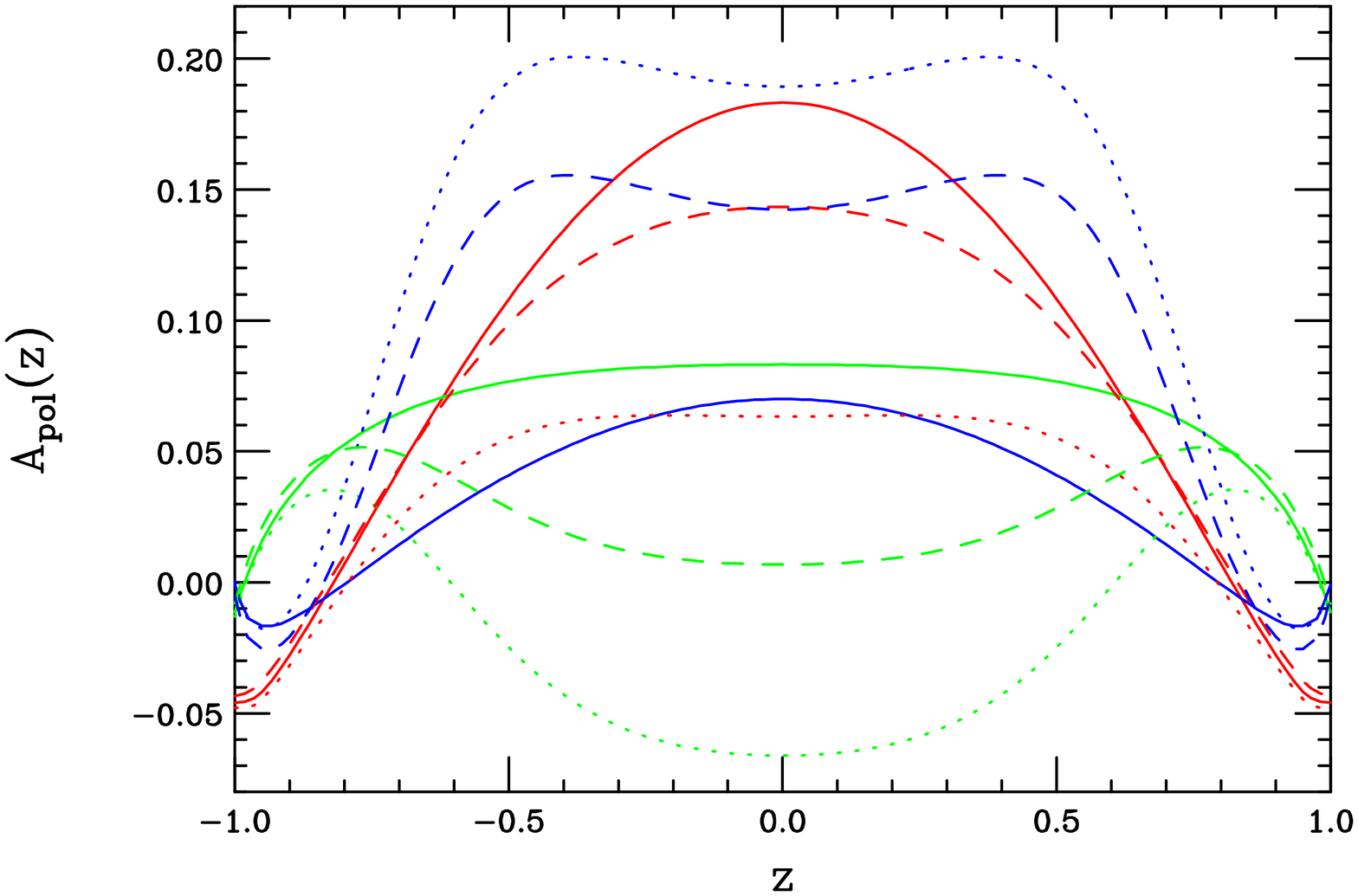,height=5.5cm,width=7cm,angle=90}}
\vspace*{0.1cm}
\caption{(Left)Fraction of LL(solid), TL+LT(dashed) and TT(dotted) $W^+W^-$ 
final states after angular cuts for the process $\gamma \gamma \to W^+W^-$ at a 
1 TeV $e^+e^-$ collider as a function of $M_H$ for either sign of $\lambda$. 
The initial state polarization in is $(-++-)$. (Right)Differential 
polarization asymmetries for $\gamma \gamma \to W^+W^-$ at a 1 TeV 
$e^+e^-$ collider for the SM(solid) as well with graviton tower exchange with 
$M_H$=2.5 TeV with $\lambda=\pm 1$(the dotted and dashed curves). We label the 
three cases shown by the first entry in the numerator in the definition of 
$A_{pol}$. Red curves (from top to bottom being the $2^{nd}$, $4^{th}$ and 
$7^{th}$) represents an initial polarization of $(++++)$, green is for the 
choice $(+++-)$(the $1^{st}$, $3^{rd}$ and 
$5^{th}$ curves) and blue is for the case $(-++-)$, (the $6^{th}$, $8^{th}$ and 
$9^{th}$ curves).}
\end{figure*}
\vspace*{4mm}

In the SM, the final state 
$W$'s are dominantly transversely polarized. Due to the nature of the spin-2 
graviton exchange, the K-K tower leads to a final state where both $W$'s are 
instead 
completely longitudinally polarized. This is shown explicitly in Fig.4 where 
we see the fraction of longitudinally polarized $W$'s falling rapidly as the 
scale $M_H$ is increased in comparison to $\sqrt s$. Thus we might expect 
that a measurement of the $W$ polarization will be a sensitive probe $M_H$. 
Other observables can be constructed out 
of the polarization dependent cross sections themselves.
For the six possible initial state polarizations one can construct 
three independent polarization asymmetries of the form 
\begin{equation}
A_{pol}(z)={{d\sigma(P_{e1},P_{l1},P_{e2},P_{l2})-d\sigma(P_{e1},P_{l1},-P_{e2},
-P_{l2})}\over {d\sigma(P_{e1},P_{l1},P_{e2},P_{l2})+d\sigma(P_{e1},P_{l1},
-P_{e2},-P_{l2})}}\,,
\end{equation}
where $z=\cos \theta$, which are shown in Fig.4. These asymmetries are not only 
functions of $z$ but are also dependent on the $W$ pair invariant mass 
$M_{WW}$ in a way that is also sensitive to graviton exchange as shown in the 
left panel of Fig.5.

\begin{figure*}
\centerline{
\psfig{figure=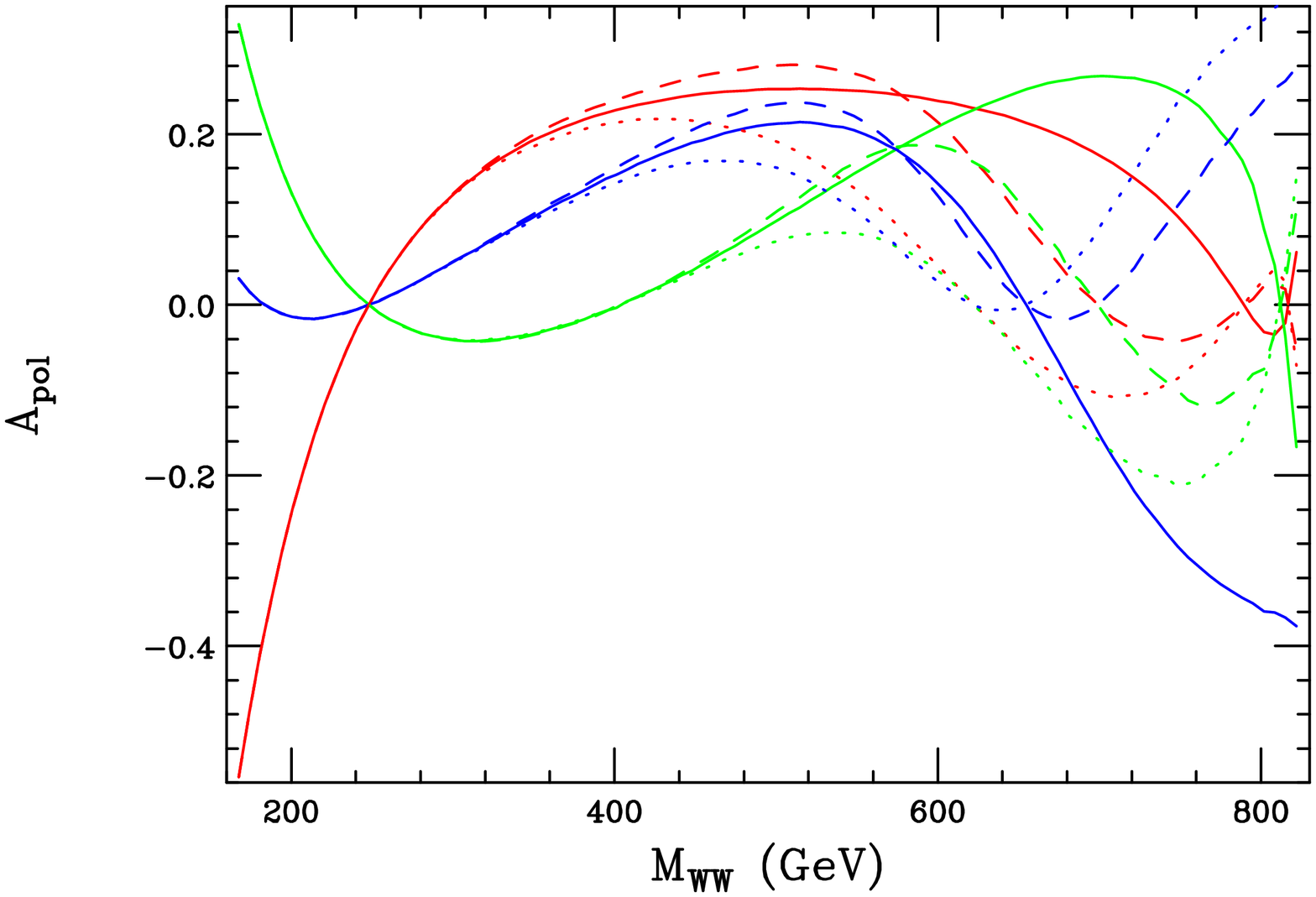,height=5.5cm,width=7cm,angle=90}
\hspace{-0.10cm}
\psfig{figure=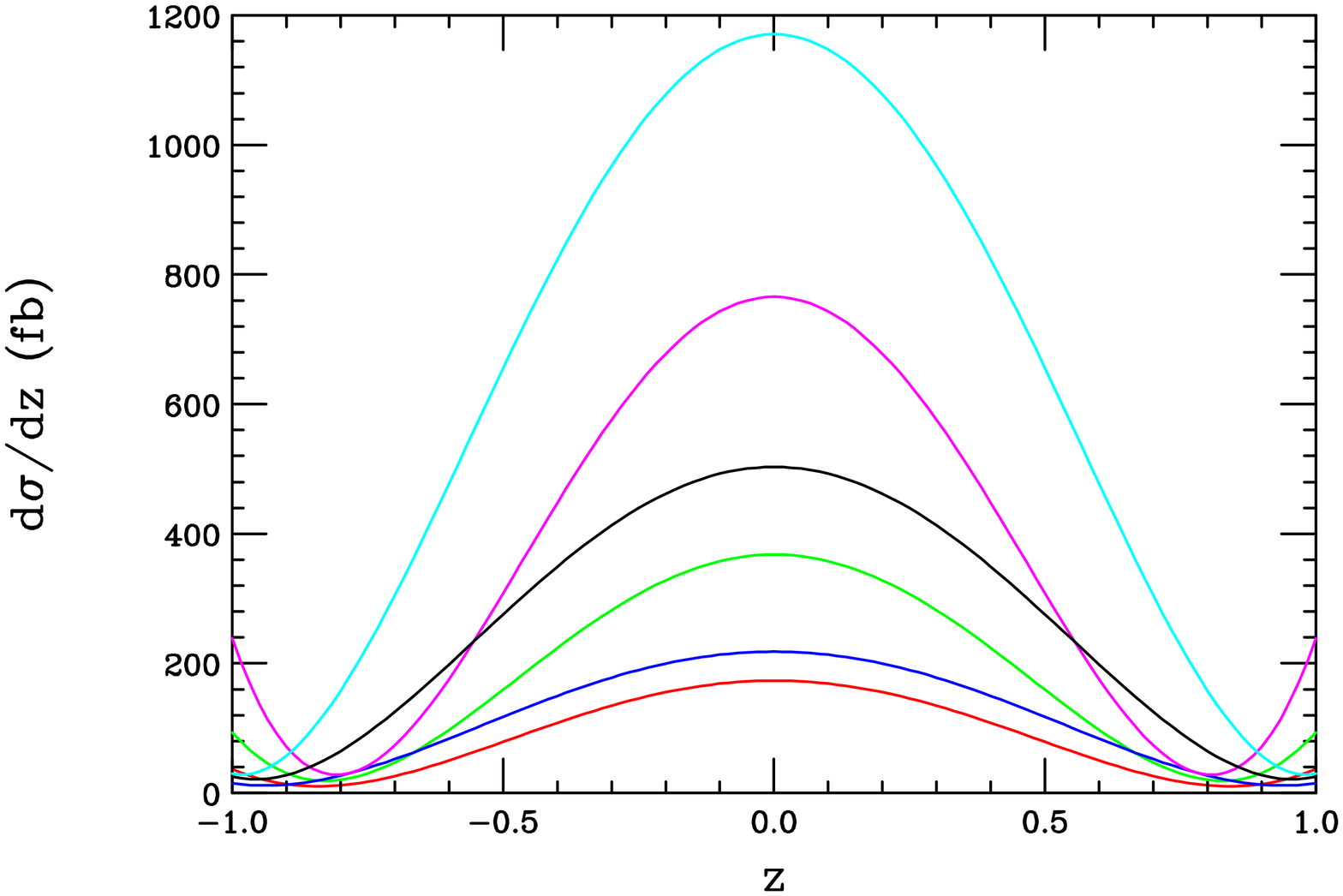,height=5.5cm,width=7cm,angle=90}}
\vspace*{0.1cm}
\caption{(Left)Integrated polarization asymmetries for 
$\gamma \gamma \to W^+W^-$ at a 1 TeV $e^+e^-$ collider as functions of the 
$WW$ invariant mass. The labels for the various curves are as in the previous 
figure and a cut of $|z|<0.8$ has been applied.
(Right)Differential cross section for $\gamma \gamma \to ZZ$ at a 1 TeV 
$e^+e^-$ collider due to the exchange 
of a K-K tower of gravitons assuming $M_H=3$ 
TeV. From top to bottom in the center of the figure the initial state 
helicities are $(-++-)$, $(+-+-)$, $(+---)$, $(+++-)$, $(++--)$, $(++++)$.}
\end{figure*}
\vspace*{4mm}

By performing a combined 
fit to the total cross sections and angular distributions, 
the $LL$ and $LT+TL$ helicity fractions for various initial state 
polarization choices and the polarization asymmetries we are able to discern 
the discovery reaches for $M_H$ as a function of the total $\gamma \gamma$ 
integrated luminosity; this is  shown in Fig.6. Here we see a search reach in 
the range of $M_H\sim 11-13\sqrt s$, which is the larger than that obtained 
from all other processes examined so far. By comparison, a combined analysis of 
the processes $e^+e^- \to f\bar f$ with the same integrated luminosity leads 
to a search reach of $\simeq 6-7\sqrt s$. 

\begin{figure*}
\centerline{
\psfig{figure=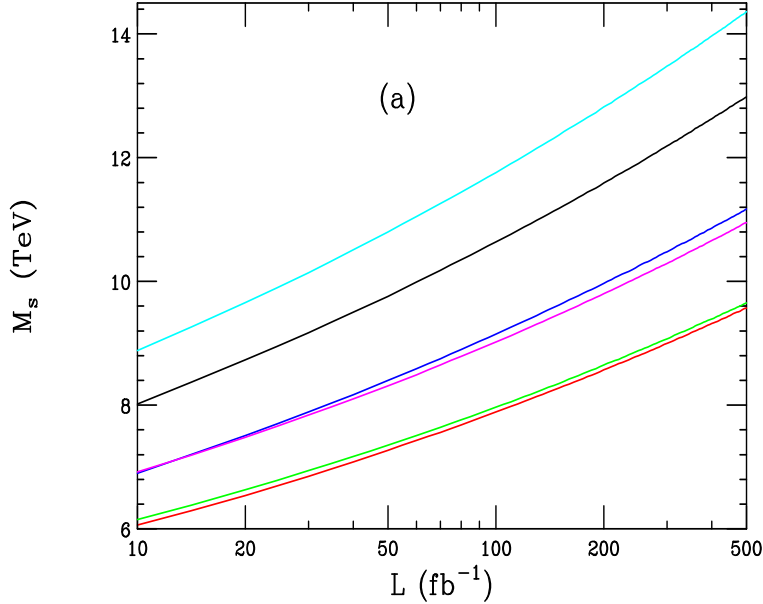,height=8cm,width=10cm,angle=90}}
\vspace*{0.1cm}
\caption[*]{$M_H$ discovery reach for the 
process $\gamma \gamma \to W^+W^-$ at a 1 TeV $e^+e^-$ collider 
as a function of the integrated luminosity 
for the different initial state polarizations assuming $\lambda=1$. 
From top to bottom on the right hand side of the figure the 
polarizations are $(-++-)$, $(+---)$, $(++--)$, $(+-+-)$, $(+---)$,  
and $(++++)$.}
\end{figure*}
\vspace*{0.4mm}

We note in passing that other $\gamma \gamma$ final states are also sensitive 
to graviton exchange, one example being $ZZ$ final state, which yield smaller 
search reaches. The right panel of Fig.5 shows 
the differential cross section due to graviton exchange for this process for 
different polarization states. These values 
are significantly larger than those arising from the SM.

\section{Conclusion}

As can be seen from the discussion above, $e^+e^-$ and $\gamma \gamma$ 
colliders are quite complementary when it comes to discovering and exploring 
new physics scenarios. In some cases, such as graviton exchange, the 
$\gamma \gamma$ reach is superior to that of other colliders.

%
\def\MPL #1 #2 #3 {Mod. Phys. Lett. {\bf#1},\ #2 (#3)}
\def\NPB #1 #2 #3 {Nucl. Phys. {\bf#1},\ #2 (#3)}
\def\PLB #1 #2 #3 {Phys. Lett. {\bf#1},\ #2 (#3)}
\def\PR #1 #2 #3 {Phys. Rep. {\bf#1},\ #2 (#3)}
\def\PRD #1 #2 #3 {Phys. Rev. {\bf#1},\ #2 (#3)}
\def\PRL #1 #2 #3 {Phys. Rev. Lett. {\bf#1},\ #2 (#3)}
\def\RMP #1 #2 #3 {Rev. Mod. Phys. {\bf#1},\ #2 (#3)}
\def\NIM #1 #2 #3 {Nuc. Inst. Meth. {\bf#1},\ #2 (#3)}
\def\ZPC #1 #2 #3 {Z. Phys. {\bf#1},\ #2 (#3)}
\def\EJPC #1 #2 #3 {E. Phys. J. {\bf#1},\ #2 (#3)}
\def\IJMP #1 #2 #3 {Int. J. Mod. Phys. {\bf#1},\ #2 (#3)}

\end{document}